\documentclass[12pt]{article}%
\usepackage[nosort]{cite}
\usepackage{graphicx}
\usepackage{multicol}
\usepackage{amsfonts}
\usepackage{amssymb}
\usepackage{amsmath}
\usepackage{heck}
\usepackage{afterpage}
\usepackage{setspace}
\usepackage{verbatim}
\usepackage{color}
\usepackage{longtable}
\usepackage{float}
\usepackage{subcaption}
\usepackage{epsfig}
\usepackage{epstopdf}
\usepackage{adjustbox}
\usepackage{tikz}
\usepackage[margin=1in]{geometry}
\usepackage{titletoc}
\usepackage{hyperref}%
\usepackage{mathrsfs}

\providecommand{\U}[1]{\protect\rule{.1in}{.1in}}
\newsavebox{\mysavebox}

\hypersetup{colorlinks,citecolor=black,filecolor=black,linkcolor=black,urlcolor=black,pdftex}
\usetikzlibrary{decorations.markings}

\numberwithin{equation}{section}

\newcommand{\be}{\begin{equation}}
\newcommand{\ee}{\end{equation}}

\usepackage{tikz}
\usetikzlibrary{arrows}
\usetikzlibrary{arrows.meta}
\usetikzlibrary{arrows,decorations.pathmorphing}
\usetikzlibrary{shapes.geometric,calc,arrows, positioning,shapes.misc,decorations.markings}
\tikzset{
  big arrow/.style={
    decoration={markings,mark=at position 1 with {\arrow[scale=2,#1]{>}}},
    postaction={decorate},
    shorten >=0.4pt},
  big arrow/.default=black}

\pgfdeclarelayer{edgelayer}
\pgfdeclarelayer{nodelayer}
\pgfsetlayers{edgelayer,nodelayer,main}
\tikzstyle{none}=[inner sep=0pt]

\tikzstyle{NodeCross}=[draw, shape=circle, cross out, inner sep=0pt, minimum size=6pt,line width=0.25mm]
\tikzstyle{Circle}=[draw, shape=circle, black, fill=black, inner sep=0pt, minimum size=6pt]
\tikzstyle{circle}=[draw, shape=circle, black, fill=black, inner sep=0pt, minimum size=16pt]
\tikzstyle{Star}=[draw, shape=star, fill=black, star points=8, inner sep=0pt, minimum size=8pt]
\tikzstyle{CircleRed}=[draw, shape=circle, black, fill=red, inner sep=0pt, minimum size=6pt]
\tikzstyle{StarP}=[draw={rgb,255: red,128; green,0; blue,128}, shape=star, fill={rgb,256: red,128; green,0; blue,128}, star points=8, inner sep=0pt, minimum size=12pt]
\tikzstyle{ShadedCircRed}=[draw=red, shape=circle, fill={rgb, 255: red,255; green,114; blue, 118}, inner sep=0pt, minimum size=80pt, line width=0.5mm, fill opacity=0.2]
\tikzstyle{ShadedCircRed2}=[draw=red, shape=circle, fill={rgb, 255: red,255; green,114; blue, 118}, inner sep=0pt, minimum size=10pt]
\tikzstyle{ShadedCircRed3}=[draw=black, shape=rectangle, fill={rgb, 255: red,255; green,114; blue, 118}, inner sep=0pt, minimum size=113pt, line width=0.25mm]
\tikzstyle{ShadedCirc}=[draw=red, shape=circle, fill=white, inner sep=0pt, minimum size=45pt,  fill opacity=1.0,  line width=0.5mm]
\tikzstyle{CircleBlue}=[draw, shape=circle, fill=blue, inner sep=0pt, minimum size=6pt]
\tikzstyle{BigCirclePurple}=[draw, shape=circle, fill={rgb,255: red,191; green,0; blue,191}, inner sep=0pt, minimum size=12pt]
\tikzstyle{CirclePurple}=[draw, shape=circle, fill={rgb,255: red,191; green,0; blue,191}, inner sep=0pt, minimum size=5pt]
\tikzstyle{EmptyCircle}=[draw, shape=circle, inner sep=0pt, minimum size=4pt]
\tikzstyle{GreenCircle}=[draw, shape=circle,  fill={rgb,255: red,80; green,200; blue,120}, inner sep=0pt, minimum size=8pt]
\tikzstyle{BrownCircle}=[draw, shape=circle,  fill={rgb,255: red,210; green,105; blue,30}, inner sep=0pt, minimum size=8pt]
\tikzstyle{CirclePurpleSmall}=[draw, shape=circle, fill={rgb,255: red,191; green,0; blue,191}, inner sep=0pt, minimum size=4pt]
\tikzstyle{BigCircleGreen}=[draw, shape=circle, fill={rgb,255: red,0; green,191; blue,0}, inner sep=0pt, minimum size=12pt]
\tikzstyle{BigCircleBlue}=[draw, shape=circle, fill={rgb,255: red,0; green,0; blue,191}, inner sep=0pt, minimum size=12pt]
\tikzstyle{BigCircleRed}=[draw, shape=circle, fill={rgb,255: red,191; green,0; blue,0}, inner sep=0pt, minimum size=12pt]
\tikzstyle{BrownCircleSmall}=[draw, shape=circle,  fill={rgb,255: red,210; green,105; blue,30}, inner sep=0pt, minimum size=6pt]

\tikzstyle{SmallCircleBrown}=[draw, shape=circle,  fill={rgb,255: red,210; green,105; blue,30}, inner sep=0pt, minimum size=5pt]

\tikzstyle{SmallCircleRed}=[draw, shape=circle, fill={rgb,255: red,191; green,0; blue,0}, inner sep=0pt, minimum size=6pt]

\tikzstyle{DashedLine}=[-, densely dashed, line width=0.25mm]
\tikzstyle{DottedLine}=[-, dotted, line width=0.25mm]
\tikzstyle{ThickLine}=[-, line width=0.25mm]
\tikzstyle{ArrowLineRight}=[-, -{Stealth[scale=1.25]}, line width=0.25mm, scale=5]
\tikzstyle{ArrowLineRed}=[-, draw={rgb,255: red,191; green,0; blue,0}, -{Stealth[scale=1.75]}, line width=0.1mm, scale=5]
\tikzstyle{RedLine}=[-, draw={rgb,255: red,191; green,0; blue,0}, fill=none, line width=0.5mm]
\tikzstyle{DashedLineThin}=[-, densely dashed, line width=0.125mm, fill=none, draw=black]
\tikzstyle{DottedRed}=[-, dotted, draw={rgb,255: red,191; green,0; blue,0}, fill=none, line width=0.25mm]
\tikzstyle{DashedRed}=[-, densely dashed, draw={rgb,255: red,191; green,0; blue,0}, fill=none, line width=0.25mm]
\tikzstyle{BlueLine}=[-, draw={rgb,255: red,0; green,0; blue,191}, fill=none, line width=0.5mm]
\tikzstyle{ArrowLineBlue}=[-, draw={rgb,255: red,0; green,0; blue,191}, -{Stealth[scale=1.75]}, line width=0.1mm, scale=5]
\tikzstyle{GreenDoubleArrow}=[<->, draw={rgb,155: red,0; green,255; blue,0},  line width= 0.5mm, scale=5]
\tikzstyle{RedDoubleArrow}=[<->, draw={rgb,255: red,255; green,0; blue,0},  line width= 0.5mm, scale=5]
\tikzstyle{BlueDottedLight}=[-, dotted, draw={rgb,255: red,0; green,0; blue,191}, fill=none, line width=0.3mm]
\tikzstyle{BrownLine}=[-, draw={rgb,255: red,210; green,105; blue,30}, fill=none, line width=0.5mm]
\tikzstyle{DottedRed}=[-, dotted, draw={rgb,255: red,191; green,0; blue,0}, fill=none, dotted, line width=0.5mm]
\tikzstyle{DottedPurple}=[-, dotted, draw={rgb,255: red,191; green,0; blue,191}, fill=none, dotted, line width=0.5mm]
\tikzstyle{BlueDottedLight}=[-, dotted, draw={rgb,255: red,0; green,0; blue,191}, fill=none, line width=0.5mm]
\tikzstyle{ArrowLinePurple}=[-, draw={rgb,255: red,191; green,0; blue,191}, -{Stealth[scale=1.75]}, line width=0.5mm, scale=5]
\tikzstyle{DashedLineGreen}=[-, densely dashed, draw={rgb,255: red,74; green,103; blue,65}, line width=0.25mm]
\tikzstyle{LineGreen}=[-, draw={rgb,255: red, 74; green,200; blue,65}, line width=0.5mm]
\tikzstyle{ArrowLineGreen}=[-, draw={rgb,255: red,0; green,191; blue,0}, -{Stealth[scale=1.75]}, line width=0.5mm, scale=5]
\tikzstyle{GreenLine}=[-, draw={rgb,255: red,0; green,191; blue,0}, fill=none, line width=0.5mm]
\tikzstyle{PurpleLine}=[-, draw={rgb,255: red,191; green,0; blue,191}, fill=none, line width=0.5mm]
\tikzstyle{PPurpleLine}=[-, draw={rgb,255: red,191; green,0; blue,191}, fill=none, line width=2.5mm]
\tikzstyle{DPurpleLine}=[-, dotted, draw={rgb,255: red,191; green,0; blue,191}, fill=none, line width=0.5mm]
\tikzstyle{SBrownLine}=[-, draw={rgb,255: red,191; green,0; blue,191}, fill=none, opacity=0.35, line width=2.5mm]
\tikzstyle{DottedBlue}=[-, dotted, draw=blue, fill=none, dotted, line width=0.5mm]

\tikzstyle{DashedPurpleLine}=[-, densely dashed, draw={rgb,255: red,191; green,0; blue,191}, fill=none, line width=0.5mm]

\tikzstyle{SmallCircleBlue}=[draw, shape=circle, fill=blue, inner sep=0pt, minimum size=5pt]
\tikzstyle{SmallCirclePurple}=[draw, shape=circle, fill={rgb,255: red,191; green,0; blue,191}, inner sep=0pt, minimum size=5pt]

\tikzset{snake it/.style={decorate, decoration=snake}}

\usetikzlibrary{decorations.markings}

\usetikzlibrary{hobby}

\tikzset{
dashstar/.style={
 dash pattern=on 5pt off 5pt,
 postaction={
  decorate,
  decoration={
   markings,
   mark=between positions 9pt and 1 step 10pt with {
     \node[color=red] {*};
   }
  }
 }
},
dashstarstar/.style={ 
 dash pattern=on 5pt off 10pt,
 postaction={
   decorate,
   decoration={
     markings,
     mark=between positions 10pt and 1
          step 15pt
           with {
            \node at (-2pt,0pt) {\pgfuseplotmark{star}};
            \node at (2pt,0pt) {\pgfuseplotmark{star}};
           }
   }
 }
}
}
\usepackage{pgfplots}
\pgfplotsset{compat=1.16}

\newcommand{\ba}{\begin{aligned}}
\newcommand{\ea}{\end{aligned}}

\begin{document}

\date{June 2024}

\title{Celestial Topology, Symmetry Theories, and \\[4mm] Evidence for a Non-SUSY D3-Brane CFT}

\institution{PENN}{\centerline{$^{1}$Department of Physics and Astronomy, University of Pennsylvania, Philadelphia, PA 19104, USA}}
\institution{PENNmath}{\centerline{$^{2}$Department of Mathematics, University of Pennsylvania, Philadelphia, PA 19104, USA}}

\authors{
Jonathan J. Heckman\worksat{\PENN,\PENNmath}\footnote{e-mail: \texttt{jheckman@sas.upenn.edu}} and
Max H\"ubner\worksat{\PENN}\footnote{e-mail: \texttt{hmax@sas.upenn.edu}}
}

\abstract{Symmetry Theories (SymThs) provide a flexible framework for analyzing the global categorical symmetries of a
$D$-dimensional QFT$_{D}$ in terms of a $(D+1)$-dimensional bulk system SymTh$_{D+1}$. In QFTs realized
via local string backgrounds, these SymThs naturally arise from
dimensional reduction of the linking boundary geometry. To track possible time dependent effects
we introduce a celestial generalization of the standard ``boundary at infinity'' of a SymTh.
As an application of these considerations we revisit large $N$ quiver gauge theories realized by spacetime filling D3-branes probing a non-supersymmetric orbifold $\mathbb{R}^6 / \Gamma$. Comparing the imprint of symmetry breaking on the celestial geometry at small and large 't Hooft coupling we find evidence for an intermediate symmetry preserving conformal fixed point.}

\maketitle

\enlargethispage{\baselineskip}

\setcounter{tocdepth}{2}


\newpage

\section{Introduction}

\begin{figure}[t!]
\begin{center}
\includegraphics[scale = 0.5, trim = {0cm 2.0cm 0cm 2.0cm}]{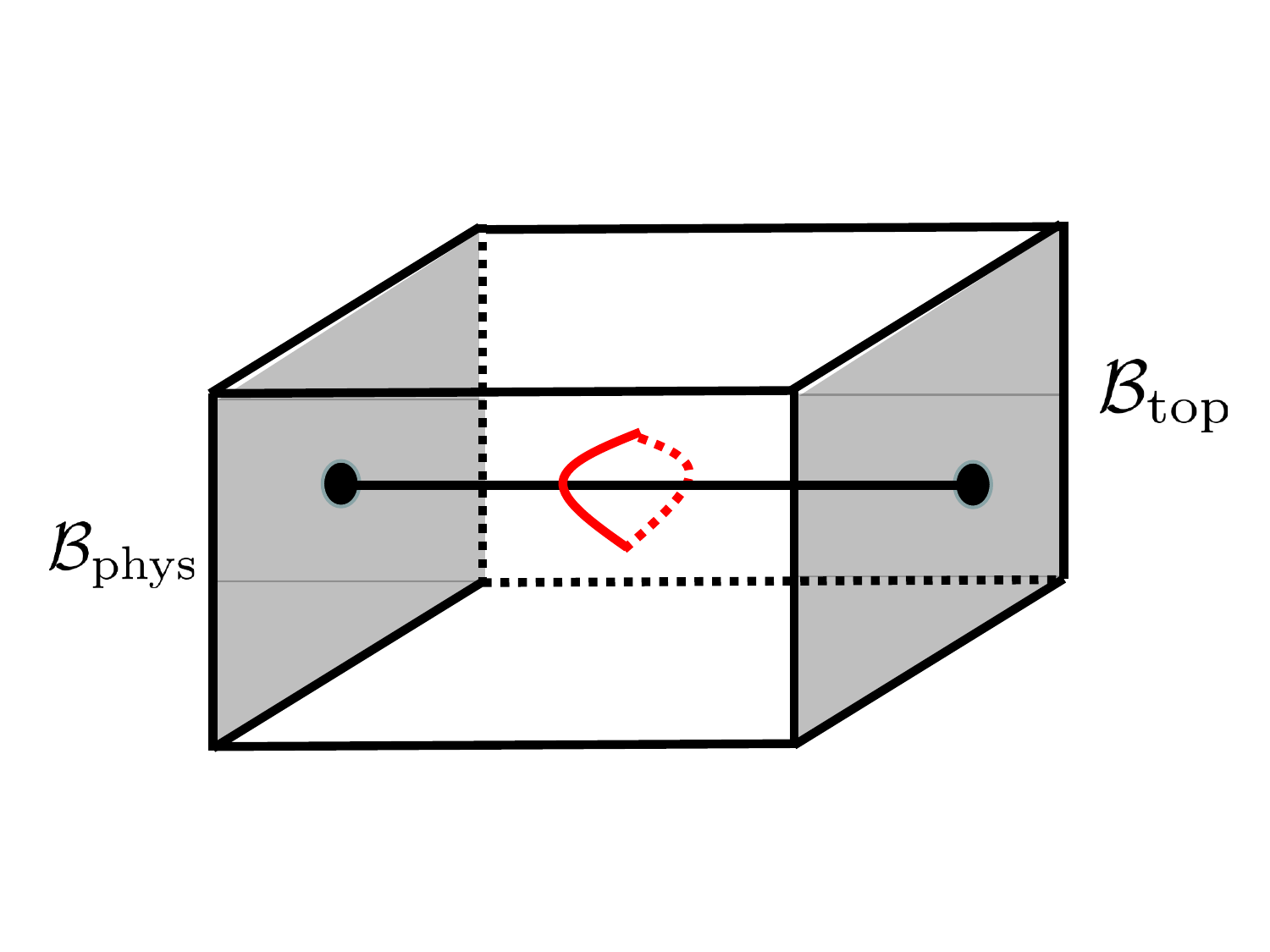}
\caption{Depiction of the standard SymTh for a finite symmetry, i.e., a SymTFT. For a $D$-dimensional QFT, the SymTFT
lives in a $(D+1)$-dimensional space with two boundaries; on one side we have a physical boundary condition $\mathcal{B}_{\mathrm{phys}}$
implementing the relative QFT, and on the other side we have a topological boundary condition $\mathcal{B}_{\mathrm{top}}$ implementing a choice of polarization. Defects stretch across the bulk direction, and symmetry operators link / intersect with such defects.}
\label{fig:SymTFT}
\end{center}
\end{figure}

Symmetries provide a helpful tool in studying the dynamics of a wide variety of quantum systems. Recent work has emphasized the topological structure underlying various generalized symmetries \cite{Gaiotto:2014kfa}.\footnote{See e.g., \cite{Cordova:2022ruw, Schafer-Nameki:2023jdn, Bhardwaj:2023kri, Luo:2023ive, Brennan:2023mmt, Shao:2023gho} for reviews.} In this setting, one needs to specify a spectrum of defects as well as topological symmetry operators which link / intersect with these defects. This linking / crossing can be implemented in terms of a higher-dimensional Symmetry Theory (SymTh).\footnote{See e.g., \cite{Reshetikhin:1991tc, Turaev:1992hq, Barrett:1993ab, Aharony:1998qu, Witten:1998wy, Fuchs:2002cm, Kirillov:2010nh, Kapustin:2010if, Kitaev:2011dxc, Fuchs:2012dt,
Freed:2012bs, Heckman:2017uxe, Freed:2018cec, Gaiotto:2020iye, Apruzzi:2021nmk, Freed:2022qnc, Kaidi:2022cpf,
Baume:2023kkf,Brennan:2024fgj, Antinucci:2024zjp, Bonetti:2024cjk, Heckman:2024oot,
Apruzzi:2024htg, DelZotto:2024tae, GarciaEtxebarria:2024fuk}.}
In the simplest situation, one decomposes the data of the absolute $D$-dimensional QFT in terms of a relative QFT,\footnote{In the sense of \cite{Freed:2012bs}.} treated as edge modes of a bulk Symmetry Theory SymTh$_{D+1}$. The specific collection of defects and global symmetries is then implemented via topological boundary conditions far from the QFT.\footnote{In the most general case there can be some subtleties with imposing purely topological boundary conditions, this will not play a role in the discussion to follow.} The overall partition function can then be visualized as an inner product of two states:
\begin{equation}\label{eq:Zpart}
Z = \langle \mathcal{B}_{\mathrm{phys}} \vert \mathcal{B}_{\mathrm{top}} \rangle,
\end{equation}
in the obvious notation.\footnote{There can be various obstructions to specific choices of boundary conditions, as captured by interaction terms of the bulk theory, which are in turn captured by anomalies of the $D$-dimensional QFT.} In cases where the symmetry is finite, the bulk SymTh is gapped and we have a Symmetry Topological Field Theory (SymTFT). In cases where the symmetry is continuous, there are various proposals for how to implement a gapped bulk.\footnote{See e.g, \cite{Brennan:2024fgj, Antinucci:2024zjp, Bonetti:2024cjk, Heckman:2024oot, Apruzzi:2024htg, DelZotto:2024tae}. These proposals tend to require a choice of regulated metric dependent boundary conditions. We comment that in general, the $(D+1)$-dimensional bulk need not be gapped \cite{Heckman:2024oot, Apruzzi:2024htg}. Furthermore, as noted in \cite{Cheesesteak}, one expects more generally a collection of nested symmetry theories which eventually filter to a gapped theory in $D+m$ dimensions with $m \geq 1$. The inner product of equation (\ref{eq:Zpart}) can then be generated from this higher-dimensional gapped theory. These subtleties will play little role in the discussion to follow.} See figure \ref{fig:SymTFT} for a depiction of the standard SymTh.

Much of this structure naturally appears in top down, i.e., stringy / M-theory realizations of QFTs.
To be concrete, consider the class of QFTs which can be engineered from asymptotically conical geometries
of the form $X = \mathrm{Cone}(\partial X)$.\footnote{One can also consider more general geometries which realize local models decoupled from $D$-dimensional gravity.}
Introducing a radial coordinate $r$ which goes from the tip of the singular cone to the conformal boundary $\partial X$,
we assume throughout that as $r \rightarrow \infty$ the metric approaches the form:
\begin{equation}
ds^2_{X} = \ dr^2 + r^2 ds^2_{\partial X},
\end{equation}
with sufficiently fast falloff as $r \rightarrow \infty$ for other radial dependent terms. This includes the asymptotic metric for quotient spaces of the form $X = \mathbb{R}^6 / \Gamma$ for $\Gamma$ a finite subgroup of $\mathrm{Spin}(6) = SU(4)$, but similar considerations hold for more general spaces. The key point for us is that dimensional reduction on $\partial X$
of the topological terms of the associated higher-dimensional action results in a topological theory on the $D$-dimensional spacetime filled by the QFT$_{D}$, as well as the interval swept out by the radial coordinate $r \in \mathbb{R}_{\geq 0}$.\footnote{See \cite{Apruzzi:2021nmk} as well as \cite{Aharony:1998qu, Heckman:2017uxe, Apruzzi:2023uma, Baume:2023kkf, GarciaEtxebarria:2024fuk}.}

A typical assumption in the stringy implementation of Symmetry Theories is that the geometry of the string background is static, i.e., independent of time. But in many cases of interest, this assumption will not hold. Examples of this sort include non-supersymmetric orbifolds with a tachyon in a twisted sector, as in references \cite{Adams:2001sv, Morrison:2004fr, Dymarsky:2005uh, Dymarsky:2005nc}. Recently reference \cite{Braeger:2024jcj} studied the higher-form symmetries for some such backgrounds. Our aim in this work will be to further study the structure of time dependent effects in these backgrounds.

While we expect our considerations to hold quite broadly, we concretely focus on the case of type IIB string theory on non-supersymmetric orbifold backgrounds of the form $\mathbb{R}^{3,1} \times \mathbb{R}^6 / \Gamma$ where $\Gamma$ is a finite subgroup of $SU(4) = \mathrm{Spin}(6)$ and we assume the group action has been chosen so that the singularity is isolated at the tip of the cone $X = \mathbb{R}^6 / \Gamma = \mathrm{Cone}(S^5 / \Gamma)$. Placing $N$ spacetime filling D3-branes at the tip of this cone leads to a non-supersymmetric 4D quiver gauge theory.\footnote{See e.g., \cite{Douglas:1996sw, Kachru:1998ys, Lawrence:1998ja} as well as \cite{Dymarsky:2005uh, Dymarsky:2005nc, Pomoni:2009joh}} One of our aims will be to study the phase structure of symmetry breaking in this gauge theory in the large $N$ limit as a function of the 't Hooft coupling $\lambda = 2 \pi g_{s} N$, which at large $N$ is a marginal parameter.\footnote{Up to parametrically controlled $1/N$ corrections.}

Explicit treatments of this system indicate the absence of a conformal fixed point both at $\lambda \ll 1$ \cite{Dymarsky:2005uh, Dymarsky:2005nc, Pomoni:2009joh} as well as at $\lambda \gg 1$ \cite{Horowitz:2007pr}, although the nature of the obstruction / instability is seemingly different in the two regimes. This leaves open the possibility that there might nevertheless be a conformal fixed point at some intermediate value of $\lambda$. By studying the causal structure of the SymTh, we argue for the existence of a critical value $\lambda = \lambda_{\ast}$ at which symmetries broken at small $\lambda$ are restored, thus providing evidence for the existence of a conformal fixed point.

\section{Causal Structures in a Symmetry Theory}

We now turn to a general discussion of causal structures in Symmetry Theories realized via top down constructions. At first sight, this sounds like a misnomer because by definition, the symmetry topological field theory (SymTFT) is suppose to be topological, i.e., independent of a choice of metric. Even in the more general setting of a SymTh which might support more general gapless excitations, the dependence on the bulk metric is expected to be quite mild. Nevertheless, there are certain aspects of the causal structure which are largely independent of the choice of a specific metric, but which can still affect the physical interpretation of states in these systems.

Along these lines, suppose we have a string background which engineers a QFT$_D$ on $D$-dimensional Minkowski space, as obtained from a
background of the form $\mathbb{R}^{D-1,1} \times X$ with metric of the form:\footnote{Of course, in actual compactifications it is sometimes necessary to multiply the $\mathbb{R}^{D-1,1}$ and $X$ directions of the metric by suitable overall warp factors, which due to the various instabilities will also be accompanied by both $r$ and $t$ dependence. However, to investigate the causal structure of the various instabilities and their impact on the celestial topology, it suffices to make the approximation used below. We emphasize that in the present context, the warp factors under considerations are those associated with probe D3-branes. As such, the asymptotic causal structure of the spacetime geometry is unaffected. This is important because we will shortly use the boundary topology to analyze symmetry breaking patterns, much as has already been carried out quite successfully in the supersymmetric setting.}
\begin{equation}
ds^2 = -dt^2 + d\overrightarrow{y} \cdot d \overrightarrow{y} + dr^2 + r^2 ds^2_{\partial X},
\end{equation}
with $\overrightarrow{y}$ the spatial coordinates of $\mathbb{R}^{D-1,1}$.

\begin{figure}[t!]
\begin{center}
\includegraphics[scale = 0.25, trim = {0cm 1.0cm 0cm 1.0cm}]{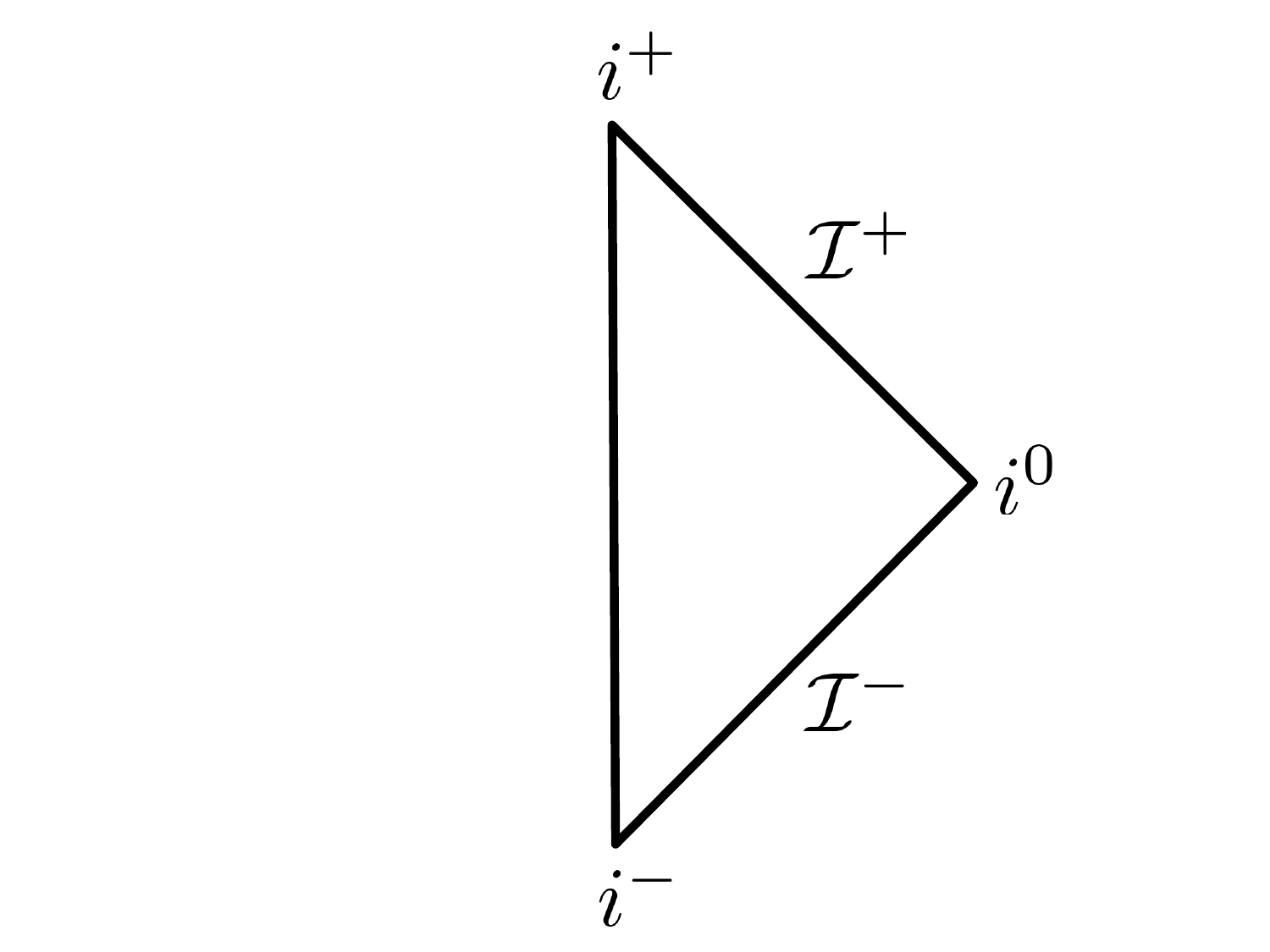}
\caption{Penrose diagram for the background $\mathbb{R}^{D-1,1} \times X$ as specified by the
2D geometry with coordinates $(T,R)$. Over each interior point of the diagram
there is a copy of $\mathbb{R}^{D-1} \times \partial X$.
}
\label{fig:PenroseSetup}
\end{center}
\end{figure}

To study the causal structure, it is convenient to introduce a Penrose diagram for the $(t,r)$ coordinates, with the remaining $\mathbb{R}^{D-1} \times \partial X$ directions treated as ``spectators''. We make the standard change of coordinates implicitly defined by $\tan(R \pm T) = r \pm t$. Since the spatial $\mathbb{R}^{D-1}$ dependence will be largely unimportant in what follows, by abuse of terminology, we shall refer to the standard $i^{\pm}$, $i^{0}$ and $\mathcal{I}^{\pm}$ (see \cite{Penrose:1962ij}) as:
\begin{align}
i^{\pm} &= \{T = \pm \pi / 2 \,\,\,\text{and} \,\,\, R = 0\} \\
i^{0} & = \{T = 0 \,\,\, \text{and} \,\,\, R = \pi / 2 \} \\
\mathcal{I}^{\pm} & = \{T = \mp (R - \pi / 2) \}.
\end{align}
We also introduce a local coordinate $\ell \in [0,1]$ for $\mathcal{I}^{+}$ along the null ray running from $i^{0}$ (at $\ell = 0$) to $i^{+}$ (at $\ell = 1$). See figure \ref{fig:PenroseSetup} for a depiction of the Penrose diagram.

Suppose now that a transition process takes place at some time in the QFT$_{D}$ located at $R = 0$. This is transmitted out to the boundary at a speed which is at most the speed of light. So, the trajectory of the signal makes it out to the boundary at some value of $\ell \in [0,1]$ which is closer to $i^{+}$ than a purely lightlike signal (see figure \ref{fig:CollisionWeakCoupling} for some examples).

Now, in the top down picture, the topological symmetry operators are captured by ``branes at infinity'' \cite{Apruzzi:2022rei, GarciaEtxebarria:2022vzq, Heckman:2022muc}, i.e., via branes wrapped on cycles of the $\partial X$
factor of $\mathcal{I}^{+}$.\footnote{In the top down approach to generalized symmetries \cite{DelZotto:2015isa, GarciaEtxebarria:2019caf, Albertini:2020mdx, Morrison:2020ool, Apruzzi:2022rei, GarciaEtxebarria:2022vzq, Heckman:2022muc, Heckman:2022xgu}, branes wrapped on (torsional) cycles of $\partial X$ generate topological symmetry operators of the QFT. One can also realize further topological operators via solitons associated with isometries of $\partial X$ \cite{Heckman:2024oot}.}

The process of symmetry breaking is thus encoded in a change of the topology of $\mathcal{I}^{+}$ before and after the collision of the signal transmission with future null infinity. Assuming we transition to a new boundary geometry $\partial X^{\prime}$, we can build a corresponding bordism between $\partial X$ and $\partial X^{\prime}$; explicit examples of this were presented in \cite{Braeger:2024jcj}. Note that all of this change is encoded in the celestial geometry of the spacetime.

Taking this into account, we observe that we can speak of a non-topological interface theory between two Symmetry Theories, much as in \cite{Baume:2023kkf, Braeger:2024jcj}. This interface lives on a ray in the $(T,R)$ directions. Further transitions in the state of the system are captured by additional rays which start at later times in the $R = 0$ slice and again propagate out to $\mathcal{I}^{+}$.

\section{Non-Supersymmetric Quiver Gauge Theory}

Let us apply these general considerations in the case of the 4D quiver gauge theory realized by type IIB string theory on the background $\mathbb{R}^{3,1} \times \mathbb{R}^{6} / \Gamma$ with $N$ D3-branes probing the orbifold singularity.
For ease of exposition we focus on the case of $\mathbb{R}^6 / \Gamma$ having an isolated singularity,
but we anticipate that many of our comments hold more broadly. Our interest will be in the phase structure of this system as a function of the
't Hooft coupling $\lambda = 2 \pi g_{s} N$.\footnote{This coupling does not run at one-loop order due to large $N$ inheritance from $\mathcal{N} =4$ Super Yang-Mills theory, and so we can meaningfully speak of tuning the parameter $\lambda$.}

The literature has primarily focused on two complementary regimes of validity, namely at $\lambda \ll 1$ where one can directly study the
QFT defined by the quiver gauge theory in perturbation theory, and $\lambda \gg 1$, where one instead has a candidate holographic dual description in terms of $AdS_5 \times S^5 / \Gamma$. In both cases, an instability has been found, though the relation (if any) between these two instabilities is less clear.

In the case of $\lambda \ll 1$, reference \cite{Dymarsky:2005uh, Dymarsky:2005nc} found that there is always a double trace operator which is not marginal in the orbifold theory. This was supported in \cite{Pomoni:2009joh} by an explicit analysis of possible radiatively generated symmetry breaking effects which indeed found that the original quiver gauge theory spontaneously breaks some of the global symmetries. It was also argued in \cite{Pomoni:2009joh} that at weak coupling there is a conformal fixed point in such systems if and only if the global symmetries are not broken.

In the case of $\lambda \gg 1$, reference \cite{Horowitz:2007pr} (see also \cite{Ooguri:2016pdq, GarciaEtxebarria:2020xsr}) found that the candidate $AdS$ dual has a charged bubble-of-nothing\footnote{See \cite{Witten:1981gj} for the original bubble of nothing instability.} type instability which eventually ``eats'' the entire $AdS$ spacetime. In the candidate CFT dual this is interpreted as an instantaneous loss of the ground state, signalling that there actually is no conformally invariant ground state.

This leaves open the possibility that at some intermediate value of $\lambda$ there may nevertheless be a conformal fixed point. We shall present evidence for this by tracking the symmetries of this system. Compared with earlier analyses, the main tool we develop is the closed string picture for symmetries which we use to put topological constraints on possible dynamics.

\subsection{The $\lambda \ll 1$ Regime}

Consider first the $\lambda \ll 1$ regime. As already mentioned, there are strong indications that in the original 4D quiver gauge theory some operators pick up a non-zero expectation value. This is a good indication that spontaneous symmetry breaking has taken place.

Indeed, this is also what we expect to happen based on the closed string picture for this background. In the case of an isolated singularity, the general picture developed in \cite{Adams:2001jb, Adams:2001sv, Harvey:2001wm, Vafa:2001ra, Morrison:2004fr} is that a tachyonic excitation in a closed string sector will condense, radiating out from the tip of the cone. To a brane probe of the singularity, this will appear as a shell which \textit{recedes away} from the brane. In particular, the distance between this shell and the brane probe will appear to grow as a function of time. Parameterizing this distance in terms of the expectation value of a scalar degree of freedom $\vert \varphi \vert$ in the worldvolume theory of the brane, we thus expect the classical value $\vert \varphi \vert$ to grow. This is in accord with what is observed in the QFT analysis where there is a radiatively induced instability which indicates that the origin of field space is unstable.

After the original tachyon has condensed, we are left with a less singular geometry which we denote by $\mathbb{R}^6 / \Gamma^{\prime}$. This pulsing off of instabilities continues until a supersymmetric singularity is eventually reached \cite{Morrison:2004fr}. For ease of exposition, let us assume that this occurs after a single pulse, and that the resulting geometry $\mathbb{R}^6 / \Gamma^{\prime}$ is a supersymmetric isolated singularity.\footnote{The topology of more general cases can be treated in a similar fashion \cite{Braeger:2024jcj}.} A D3-brane probe of $\mathbb{R}^6 / \Gamma^{\prime}$ is then a supersymmetric quiver gauge theory, and this is expected to flow to a 4D $\mathcal{N} = 1$ superconformal field theory (SCFT). The SymTh for this system can be worked out by dimensional reduction on the boundary generalized lens space $S^5 / \Gamma^{\prime}$, as performed in \cite{Heckman:2022xgu}, in line with the reduction procedure of reference \cite{Apruzzi:2021nmk}.

\begin{figure}[t!]
\begin{center}
\includegraphics[scale = 0.5, trim = {0cm 0.0cm 0cm 1.0cm}]{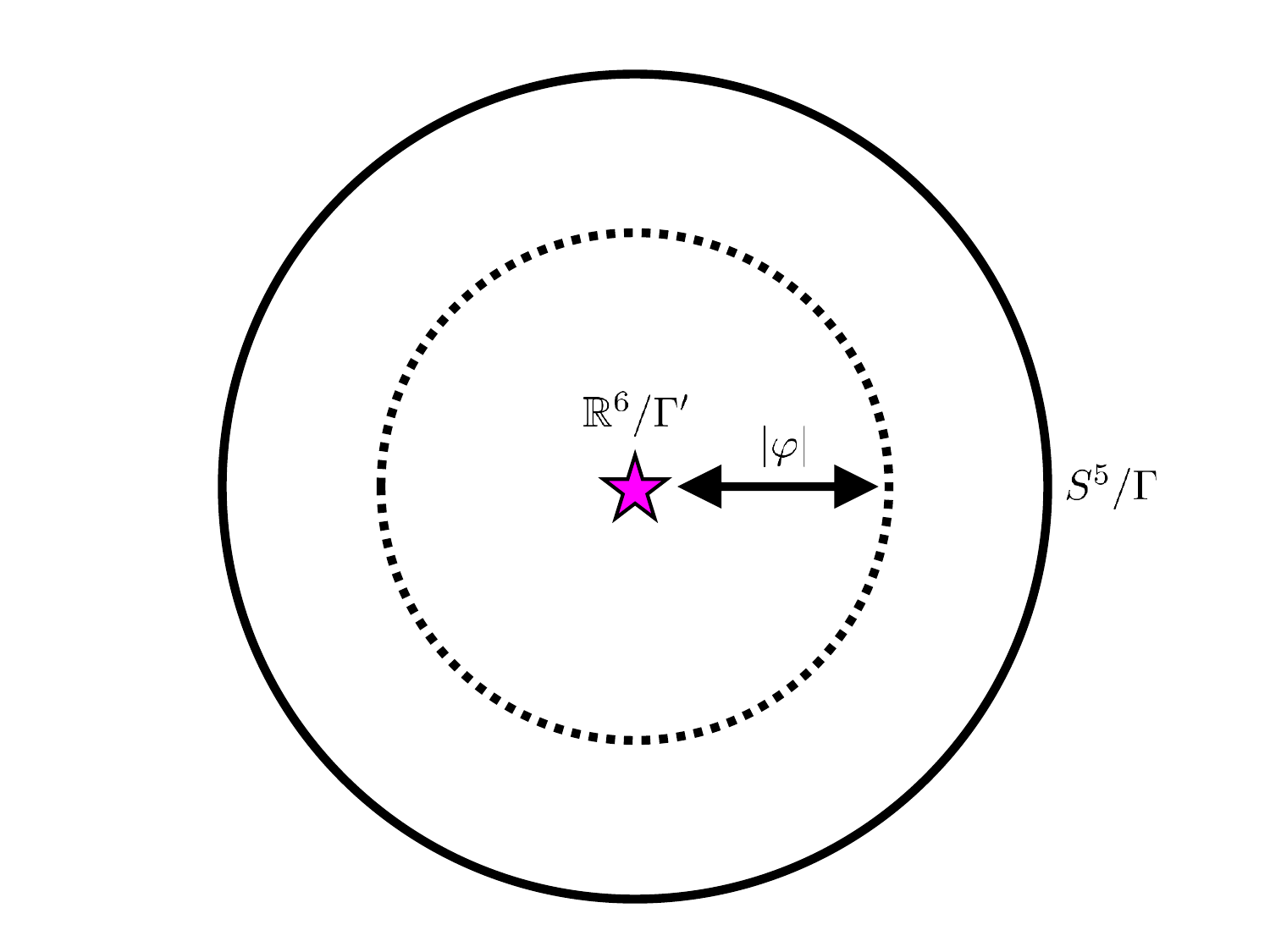}
\caption{Depiction of a D3-brane probe of the supersymmetric singularity $\mathbb{R}^6 / \Gamma^{\prime}$ after condensation of the twisted sector tachyon in the $\mathbb{R}^6 / \Gamma$ geometry. The separation between the shell and the D3-brane is captured by a scalar $\varphi$; there is a corresponding effective potential which separates the unstable $\mathbb{R}^6 / \Gamma$ region from the stable (and supersymmetric) $\mathbb{R}^6 / \Gamma^{\prime}$ region.}
\label{fig:TachyonShells}
\end{center}
\end{figure}

\begin{figure}[t!]
\begin{center}
\includegraphics[scale = 0.60, trim = {0cm 4.0cm 0cm 2.0cm}]{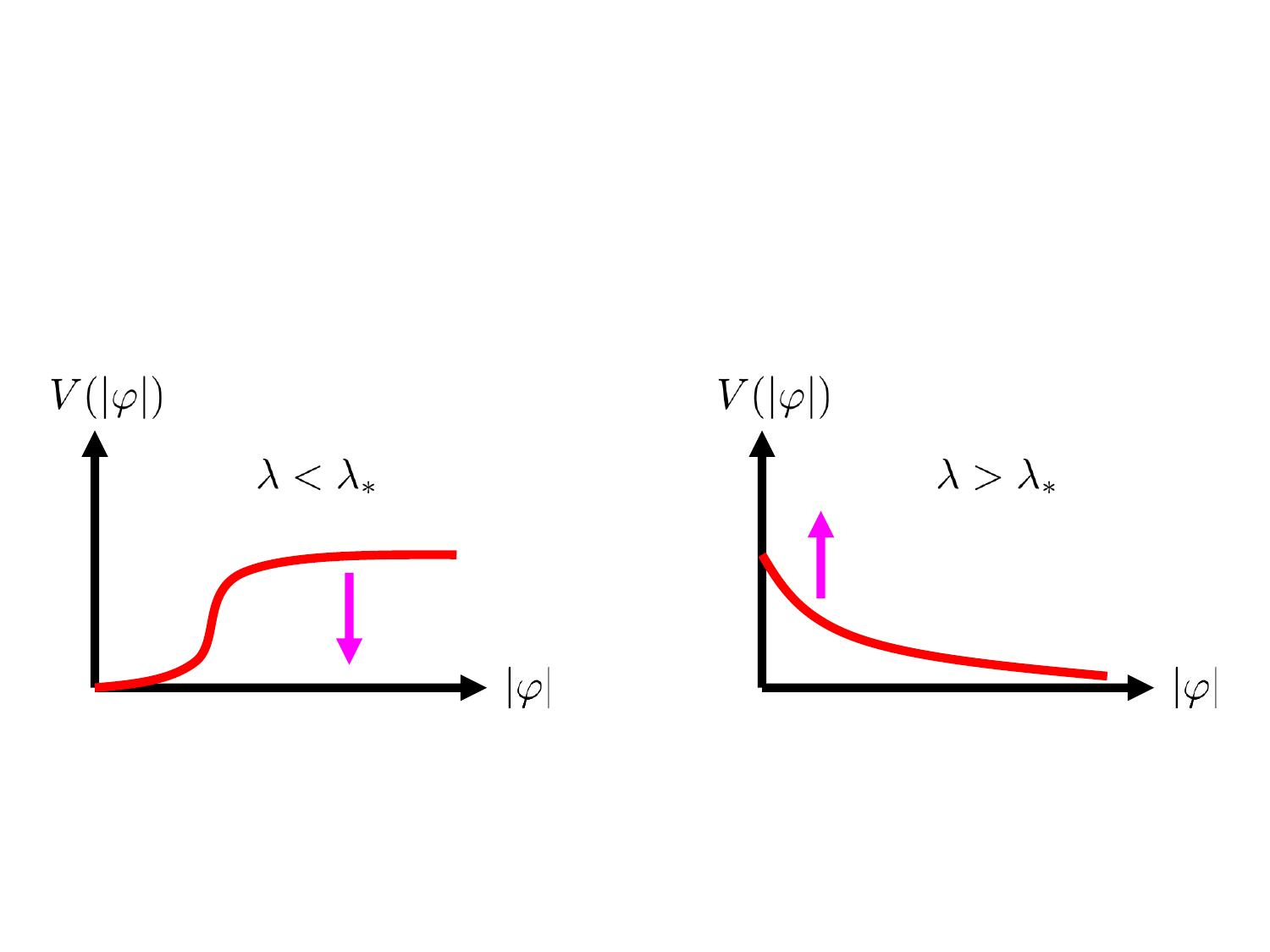}
\caption{Depiction of the proposed effective potential for a scalar $\varphi$ which parameterizes the distance of the D3-brane from the tip of the cone. At small 't Hooft coupling, there is a local minimum at the origin of field space (lefthand side). Increasing $\lambda$ lowers the barrier, until eventually some critical value of $\lambda_{\ast}$ is passed, at which point the minimum is destroyed altogether (righthand side).
This figure is meant as a qualitative characterization, a more complete calculation remains an outstanding open problem.}
\label{fig:PotentialInterpolator}
\end{center}
\end{figure}

This provides a natural candidate for the endpoint of the symmetry breaking process. The true ground state of the system is the supersymmetric quiver gauge theory obtained from D3-branes probing $\mathbb{R}^6 / \Gamma^{\prime}$ and prior to flowing to the deep infrared, there is an effective potential where displacement from the origin of $\mathbb{R}^6 / \Gamma^{\prime}$ corresponds to motion up the effective potential. From the perspective of the closed string background, the D3-brane perceives a shell of energy moving away from it (see figure \ref{fig:TachyonShells}). See figure \ref{fig:PotentialInterpolator} for a sketch of the proposed effective potential at small $\lambda$. Let us emphasize that this potential is compatible with various topological constraints, but a detailed calculation of its precise shape remains an outstanding open problem.

At this point, let us address one particularly subtle aspect of our analysis so far; we are not simply considering an excited state of the QFT and watching its trajectory as a function of time. In that case we would observe an oscillation where the fields move to the minimum of the effective potential and then move back up. Rather, what is happening is that as we roll from the initial excited state down to the minimum of the effective potential, some energy is being radiated away as closed string modes; this is simply the tachyonic shell moving away from the D3-brane. Consequently, if we continue to work purely about the ground state specified by the $\mathbb{R}^6 / \Gamma^{\prime}$ geometry we will instead perceive a ``vertical wall'' in the associated SymTh. This is the point of view adopted in \cite{Braeger:2024jcj}; here we are focusing on how the closed string dynamics is encoded in the D3-brane worldvolume theory. Again, the point here is that we are taking at face value the UV non-supersymmetric quiver gauge theory description and simply tracking what happens at long distance scales.

To give a more complete characterization of the SymTh derived from this picture,
let us now turn to some additional features of its causal structure.
Following references \cite{Adams:2001jb, Adams:2001sv, Horowitz:2007pr},
we know that the tachyonic mass squared of the original instability is of the general form:
\begin{equation}
\alpha^{\prime} M^2 = \kappa_{\Gamma} \lambda^{1/2} - \vert \xi_{\Gamma} \vert
\end{equation}
where $\kappa_{\Gamma}$ and $\xi_{\Gamma}$ are ``order one'' parameters which depend on the details of the quotient space. The second term is the one calculated from the worldsheet zero-point energy contributions of the (tachyonic) twisted sector(s), and the first term originates from a finite tension closed string stretched over a finite radius $S^5 / \Gamma$. For example \cite{Adams:2001jb, Adams:2001sv, Horowitz:2007pr},
in the case of $\mathbb{R}^{6} / \Gamma$ with $\Gamma = \mathbb{Z}_k$ and group action $(1,1,1)$ on holomorphic coordinates $\mathbb{C}^3 \simeq \mathbb{R}^6$, one has $\kappa_{\Gamma} = 1/k^2$ and $\xi_{\Gamma} = 2(3-k)/k$.

\begin{figure}[t!]
\begin{center}
\includegraphics[scale = 0.5, trim = {0cm 1.0cm 0cm 1.0cm}]{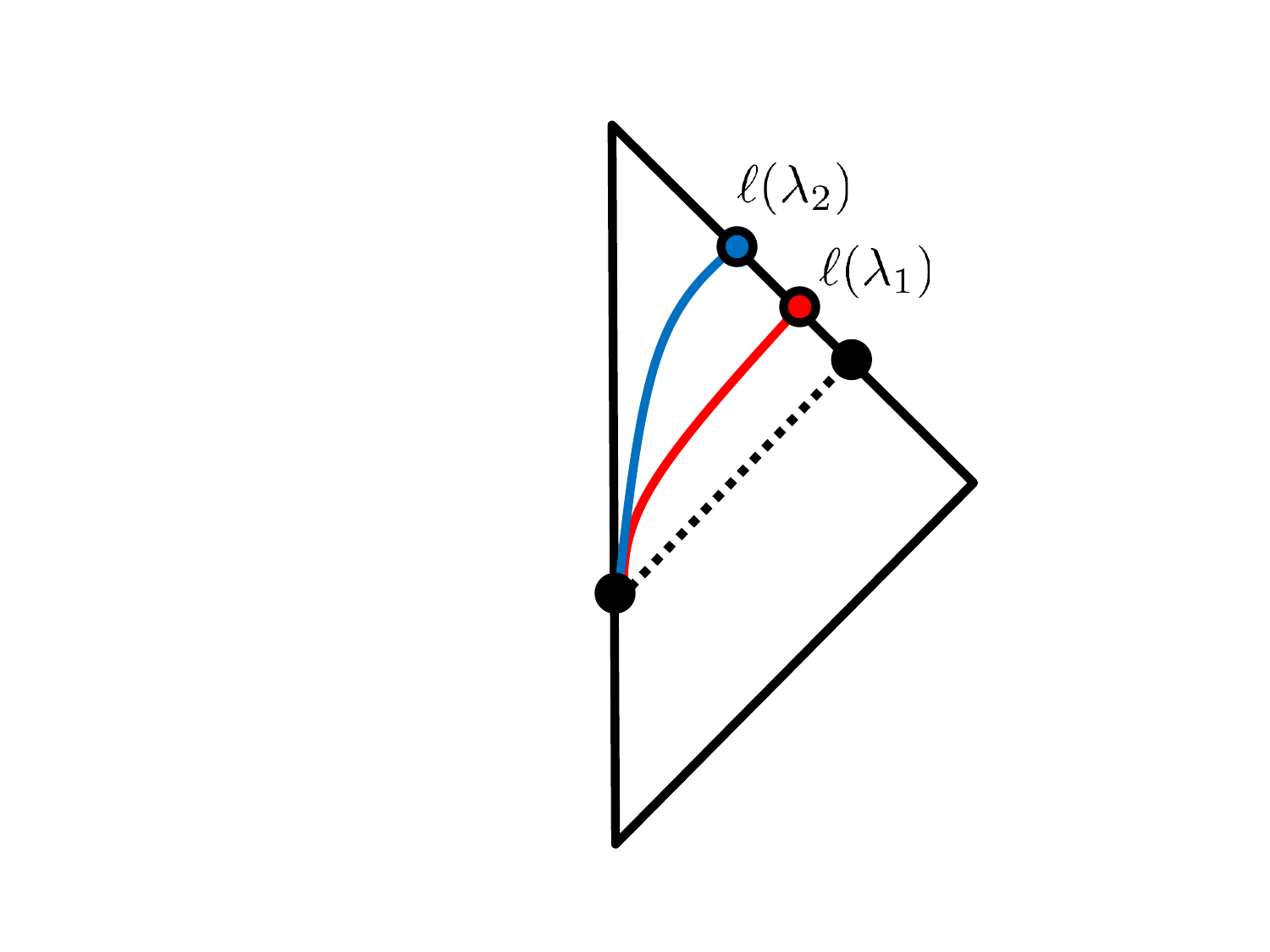}
\caption{Depiction of how the instability of the QFT$_D$ is transmitted out to $\mathcal{I}^{+}$ as a function of $\lambda$. Initially,
the signal proceeds at slower than the speed of light, but eventually asymptotes to near the speed of light.
For $\lambda_{2} > \lambda_1$ we have $\ell(\lambda_2) > \ell(\lambda_1)$. For reference we have also indicated the null ray associated with a lightlike signal (dashed black).}
\label{fig:CollisionWeakCoupling}
\end{center}
\end{figure}

Observe that as we increase $\lambda$ (but still keeping $\lambda \ll 1$), the magnitude of $M^2$ decreases. Thus, the initial speed of the transmission in the SymTh is slower than the speed of light.\footnote{``Tachyonic'' simply means we have an unstable configuration when $M^2 < 0$. Nothing is moving faster than the speed of light. Increasing the magnitude of $M^2$ in this regime just makes the effective potential steeper and steeper. Conversely, decreasing the magnitude of $M^2$ makes the effective potential more shallow.} On the other hand, at late times, we know that the instability is carried out by a massless metric and dilaton pulse, and so asymptotically approaches the speed of light. Putting these facts together, we conclude that in the Penrose diagram, the eventual collision of the trajectory in the $(T,R)$ directions with $\mathcal{I}^{+}$ continues to move closer to $i^{+}$ as we increase $\lambda$ (see figure \ref{fig:CollisionWeakCoupling}). In the interpolating region between $S^5 / \Gamma$ and $S^5 / \Gamma^{\prime}$ there is a bordism which connects the two pieces of $\mathcal{I}^{+}$.\footnote{It is natural to ask whether having $\lambda \neq 0$ simply leads to a runaway potential, i.e., the ground state is simply destroyed altogether. This is incompatible with the closed string picture. Indeed, from \cite{Adams:2001sv, Morrison:2004fr}, we know that the probe D3-branes essentially stay put and, after the tachyonic shell has smoothed away the non-supersymmetric parts of the singularity, eventually the system settles in a supersymmetric geometry probed by the D3-branes. In field theory terms, this means that the system eventually settles in a supersymmetric ground state. We emphasize that this is the behavior in the regime $\lambda \ll 1$.}

Of course, it is tempting to extrapolate this result to larger values of $\lambda$,
but then we are far outside the regime of perturbation theory. To gain some purchase,
we next turn to the holographic dual description of $\lambda \gg 1$
and then use topological considerations to constrain possible interpolations.

\subsection{The $\lambda \gg 1$ Regime}

Let us now turn to the $\lambda \gg 1$ regime. Based on experience with supersymmetric examples, it is tempting to simply take the near horizon limit of the stack of $N$ D3-branes, and in so doing reach a closed string background of the form $AdS_5 \times S^5 / \Gamma$.
Caution is warranted, however, because there is an order of limits subtlety having to do with the timescale for possible instabilities being included / excluded from the near horizon geometry. For example, if we consider the late time geometry where $\mathbb{R}^6 / \Gamma$ has already decayed to $\mathbb{R}^6 / \Gamma^{\prime}$, then in a suitable scaling limit we would instead get $AdS_5 \times S^5 / \Gamma^{\prime}$, a supersymmetric background.

Nevertheless, taking the near horizon geometry at face value, we can proceed as in \cite{Horowitz:2007pr}. Ultimately, we will be interested in the asymptotically flat setup, however it is useful to consider this putative $AdS$ dual to characterize the instabilities that might occur away from such a near horizon limit. In this near horizon geometry the appearance of an $S^5 / \Gamma$ factor means that in the ``branes at infinity'' picture we have the corresponding topological symmetry operators induced directly from the topology of this geometry. Moreover, the original tachyonic mode of the twisted sector has been stretched to a large value, and is no longer present \cite{Adams:2001jb, Adams:2001sv, Horowitz:2007pr}. On the other hand, the appearance of a non-perturbative bubble destroys the solution at finite proper time; the putative CFT dual seems to have disappeared altogether!

Somehow, the shape of the effective potential found at $\lambda \ll 1$ has changed. First of all, we can exclude the possibility that the $\mathbb{R}^6 / \Gamma^{\prime}$ vacuum found at weak coupling has become metastable; if that were the case the bubble would appear to have broken some of the spatial symmetries of $\mathbb{R}^{D-1,1}$ spacetime supporting the QFT, as would occur in a transition between vacua \cite{Coleman:1977py, Callan:1977pt}. This is not what is found in \cite{Horowitz:2007pr}.\footnote{As supporting evidence for this interpretation, observe that there is no stable Euclidean D3-brane instanton configuration which interpolates from $r = 0$ to some large value of $r$ (there is no relative 1-cycle in $H_1(X,\partial X)$).} Rather, the effective potential has simply tilted over; there is now a runaway direction rather than a minimum of the effective potential (see figure \ref{fig:PotentialInterpolator}). In the scaling limit associated with a CFT, this runaway is ``instantaneous''.

\begin{figure}[t!]
\begin{center}
\includegraphics[scale = 0.5, trim = {0cm 1.0cm 0cm 2.0cm}]{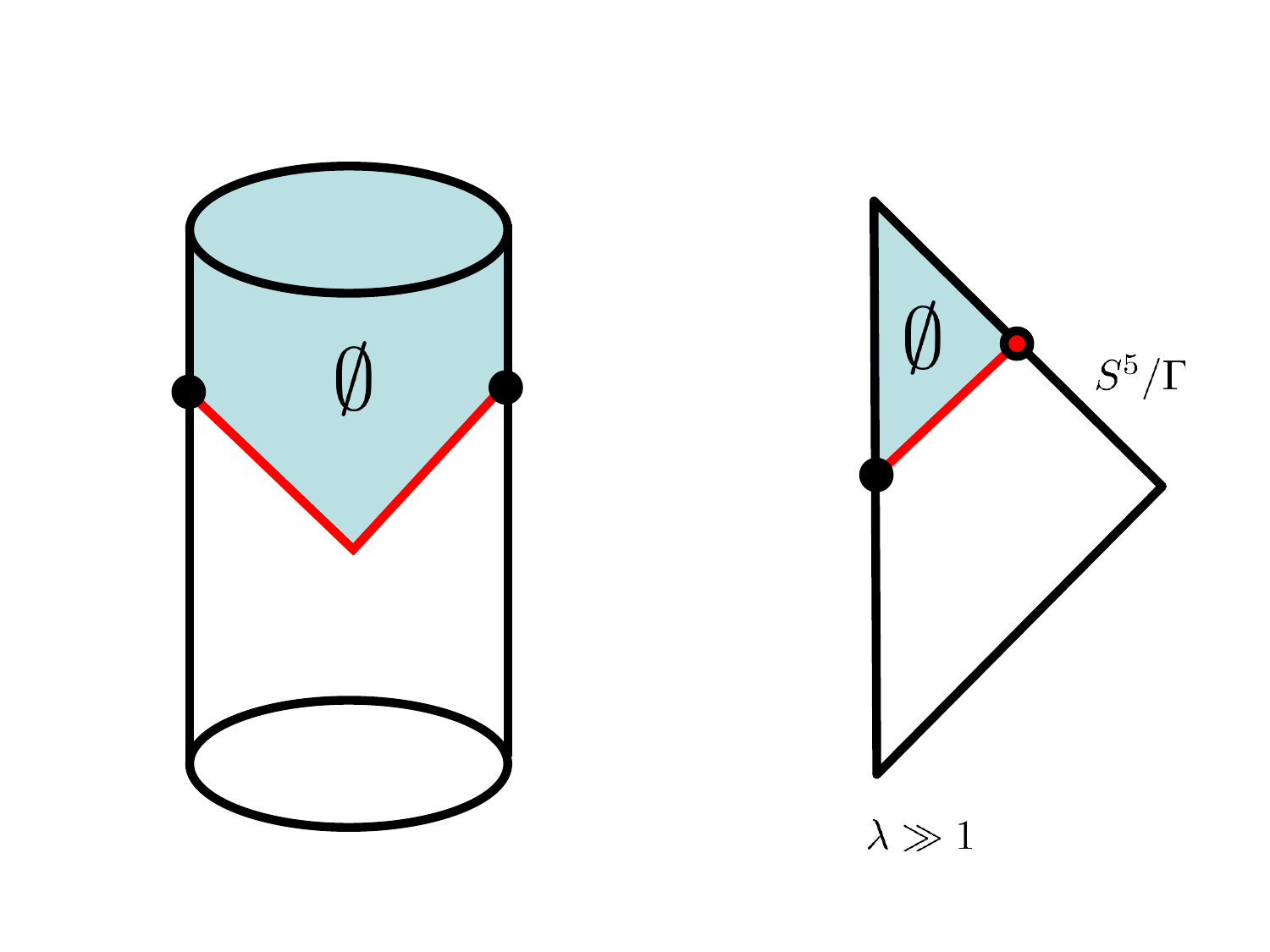}
\caption{Depiction of the bubble-of-nothing type instability in $AdS_5 \times S^5 / \Gamma$ (left) and its characterization in the SymTh for the QFT localized at $r = 0$ (right). On the lefthand side we show the Penrose diagram for global $AdS_5$, i.e., an infinitely long cylinder in the time direction, but with the instability nucleated at some finite time, as in \cite{Horowitz:2007pr}. On the righthand side we show the Penrose diagram for asymptotic flat space, i.e., away from the near horizon limit of the D3-branes. The instability found in the near horizon geometry continues to propagate out, producing on $\mathcal{I}^+$ a celestial bordism $S^5 / \Gamma \rightarrow \emptyset$.}
\label{fig:AdSGlueSymTFTnew}
\end{center}
\end{figure}

Let us now piece this together with the SymTh picture. Far away from the stack of D3-branes, we still have asymptotically
flat space, so we can ask how an instability is transmitted to $\mathcal{I}^{+}$ as it moves out of the $AdS$ throat.
To set conventions, introduce coordinates and metric for global $AdS_5$ with radius of curvature $L$:
\begin{equation}
ds^2 = \frac{L^{2}}{\cos^2 \chi} \left(-dt^2 + d \chi^2 + \sin^2 \chi d \Omega^2_{3} \right),
\end{equation}
with $0 \leq \chi \leq \pi / 2$. The SymTh amounts to a thin sliver sitting just beyond
the conformal boundary at $\chi = \pi / 2$ \cite{Heckman:2024oot}.\footnote{In the original SymTh,
the physical boundary condition on a four-dimensional boundary has been extended to the five-dimensional
gravitational bulk.} We extend this sliver into the bulk to glue the $AdS$ throat to the rest of the geometry.
Introduce a radial cutoff at
\begin{equation}
\chi_{\mathrm{max}} = \frac{\pi}{2} - \varepsilon.
\end{equation}
For $\chi > \chi_{\mathrm{max}}$ we can choose a metric which then asymptotes to flat space (far from the D3-branes).
So, the bubble nucleated at some finite time in the bulk $AdS$ expands out of the throat at the speed of light.
The bubble continues to proceed outwards, colliding with $\mathcal{I}^{+}$.
After the bubble hits $\mathcal{I}^{+}$, asymptotic null infinity cease to exist, i.e., we have a bordism of $S^5 / \Gamma$ to ``nothing''.
Clearly, the celestial topology for the $S^5 / \Gamma \rightarrow \emptyset$ transition at $\lambda \gg 1$
is different from that of the $S^5 / \Gamma \rightarrow S^5 / \Gamma^{\prime}$ transition found at $\lambda \ll 1$.
See figure \ref{fig:AdSGlueSymTFTnew} for a depiction.

\subsection{Symmetry Restoration / Evidence for a Non-SUSY CFT}

Summarizing the story so far, we have argued that in the $\lambda \ll 1$ regime there is a closed string tachyon and that as we increase $\lambda$ (while keeping it small!) the collision of the non-topological interface with $\mathcal{I}^+$ occurs closer and closer to $i^+$. On the other hand, once we increase to very large values of $\lambda \gg 1$, $i^+$ seems to have disappeared altogether, since part of $\mathcal{I}^{+}$ has now been eaten up by a bubble-of-nothing type instability.

What happens in the interpolating range of values for $\lambda$? Our claim is that varying $\lambda$ inevitably leads to a transition which can be pushed all the way to $i^+$. To see this, introduce $\ell$ a local coordinate on the null ray extending from
$i^0$ (i.e., $\ell = 0$) to $i^+$ (i.e., $\ell = 1$), we denote by $\ell(\lambda)$ the value of $\ell$ at which the instability of the SymTh intersects $\mathcal{I}^{+}$. A direct calculation of $\ell(\lambda)$ appears out of reach, but we can nevertheless constrain it based on the topology of the celestial geometry. Note first that in the regime $\lambda \ll 1$, the function $\ell(\lambda)$ is monotonically increasing. On the other hand, it somehow must ``turn around'' since eventually near $\lambda \gg 1$ the bubble-of-nothing type instability quickly eats the whole spacetime. There is some critical value of $\ell_{\mathrm{closed}}$ beyond which we no longer have an $S^5 / \Gamma \rightarrow S^5 / \Gamma^{\prime}$ transition at the boundary, i.e., we cease to have the original closed string tachyon. Likewise, there is some critical value of $\ell_{\mathrm{bubble}}$ which signals the first onset of the bubble-of-nothing type instability, i.e., we instead have a transition $S^5 / \Gamma \rightarrow \emptyset$.

The only option which appears to fit with the causal structure of the SymTh is to let $\ell(\lambda)$ eventually reach $\ell = 1$. Once this occurs, it can switch directions and proceed back down to smaller values of $\ell$. After switching directions, we can gradually start eating up more and more of $\mathcal{I}^+$, eventually reaching the configuration at $\lambda \gg 1$ where half of $\mathcal{I}^{+}$ is simply gone. Indeed, suppose to the contrary that instead the turnaround happens at some other value $\ell_{\mathrm{turn}} < 1$. We then face an immediate problem with what to do with the topology of the $\mathbb{R}^{3} \times S^5 / \Gamma^{\prime}$ region for values of $\ell > \ell_{\mathrm{turn}}$; there is no physical process which can suddenly appear to ``destroy'' this region of the spacetime.

\begin{figure}[t!]
\begin{center}
\includegraphics[scale = 0.5, trim = {0cm 1.0cm 0cm 2.0cm}]{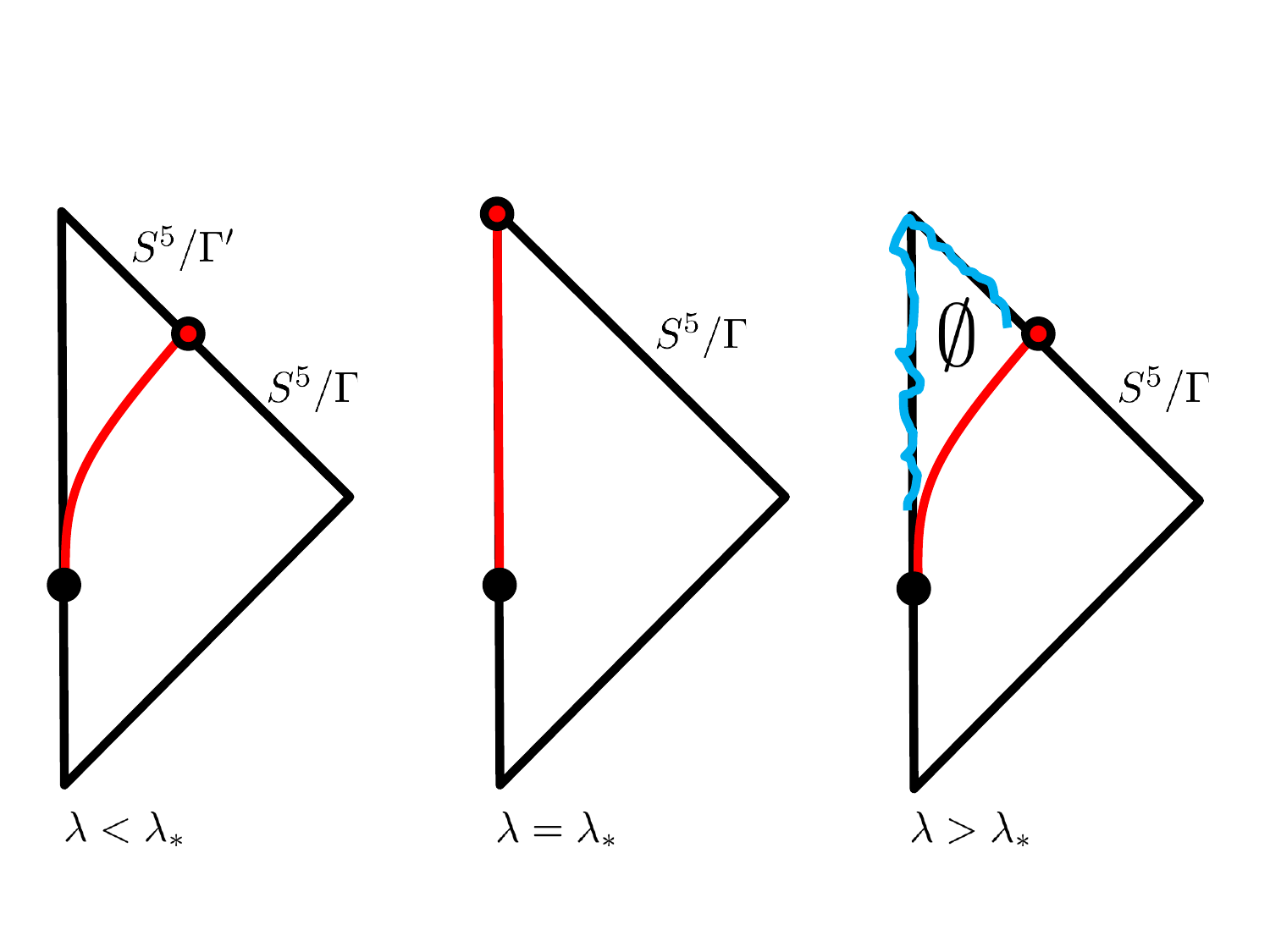}
\caption{Depiction of how the instability of the QFT$_D$ is transmitted out to $\mathcal{I}^{+}$ as a function of $\lambda$. For $\lambda < \lambda_{\ast}$, we have a bordism which connects $S^5 / \Gamma$ and $S^5 / \Gamma^{\prime}$. For $\lambda > \lambda_{\ast}$, there is instead an instability which eats up the spacetime of the SymTh. At a critical value of $\lambda = \lambda_{\ast}$ the $S^5 / \Gamma$ is retained throughout all of $\mathcal{I}^+$.}
\label{fig:PenroseInterpolator}
\end{center}
\end{figure}

Putting together these constraints on the topology of $\mathcal{I}^{+}$ and its collision with the interface of the SymTh, we conclude that $\ell_{\mathrm{closed}} = \ell_{\mathrm{bubble}} = 1$. Observe that at this point, there is a precarious balancing between the different possible instabilities. Nevertheless, what is clear is that the boundary topology of $\mathcal{I}^+$ retains the full $S^5 / \Gamma$, a good indication that the symmetries have \textit{not} been spontaneously broken at all. See figure \ref{fig:PenroseInterpolator} for a depiction of how the different instabilities are transmitted to $\mathcal{I}^{+}$.

This also suggests the following picture for the effective potential (see figure \ref{fig:PotentialInterpolator}). Starting with the original minimum which preserves the $S^5 / \Gamma^{\prime}$ symmetries, the maximum associated with the $S^5 / \Gamma$ instability continues to lower until the whole effective potential ``flattens out''. Indeed, in the limit of flowing to the deep infrared we expect that all local ``wiggles'' in the effective potential either shrink to zero size or grow to be enormous. Once we reach the critical symmetry restoring value
$\lambda = \lambda_{\ast}$, we observe that any further increase destroys the ground state itself; the minimum lifts off of zero and a rolling / runaway solution appears. The lifting off begins rather gradually though in the regime $\lambda \gg 1$ the potential is quite steep.

Summarizing, we have argued based on the topology of the celestial geometry that there is a critical value of $\lambda_{\ast}$ at which the symmetries associated with the boundary topology $S^5 / \Gamma$ are indeed preserved. We interpret this as evidence for the existence of a non-supersymmetric conformal fixed point.

Let us further emphasize, that indeed in the transition from large to small values of $\lambda$, across $\lambda_*$, the potentials displayed in figure \ref{fig:PotentialInterpolator} are guaranteed to have distinct large field profile, as characteristic for a second order phase transition. For example, a potential with ever rolling solution at small 't Hooft coupling is excluded by the perturbative analysis at small $\lambda$. There, following \cite{Braeger:2024jcj}, we know that after the background relaxes to a stable configuration we are left with a D3-brane probing a supersymmetric geometry, i.e., we are guaranteed to settle into a stable vacuum.

The considerations here have focused on topological properties of the celestial geometry. It would be interesting to connect these considerations
more directly to celestial holography, though we leave the exploration of this speculation for future endeavors.\footnote{See e.g., \cite{Pasterski:2021raf} for a recent review of celestial holography.}

Finally, it is intriguing that consistency of the picture also implies the existence of a nearly flat effective potential in the $\lambda \sim \lambda_{\ast}$ regime. This is likely of interest in engineering slow roll / nearly flat quintessence-like potentials in string theory. It would also be interesting to study potential constraints arising from coupling to gravity, as in the Swampland program (see e.g., \cite{Obied:2018sgi}).


\section*{Acknowledgements}

We thank N. Braeger, V. Chakrabhavi, I.R. Klebanov, C. Murdia, and E. Pomoni
for helpful correspondence and discussions. We thank C. Murdia for comments on an earlier draft.
The work of JJH is supported by DOE (HEP) Award
DE-SC0013528. The work of JJH and MH is supported in part by BSF grant 2022100.
The work of MH is also supported by the Simons Foundation Collaboration grant \#724069 on
\textquotedblleft Special Holonomy in Geometry, Analysis and
Physics\textquotedblright.


\bibliographystyle{utphys}
\bibliography{TimeSymTFT}

\end{document}